\documentclass[%
reprint,
aps,
prl,
amsmath,
twocolumn,
amssymb,
floatfixng,
showpacs,
superscriptaddress,
footinbib,
multicol
]{revtex4-2}
\usepackage{graphicx}
\usepackage{epstopdf}
\usepackage{dcolumn}
\usepackage{bm}
\usepackage{color}
\usepackage{xcolor}
\usepackage{amsmath}
\usepackage{braket}
\usepackage{amsfonts}
\usepackage{soul}
\usepackage{url}
\usepackage{array,tabularx}
\usepackage{graphics}
\usepackage{times}
\usepackage{subfigure}

\usepackage[export]{adjustbox}
\usepackage[colorlinks=true,linktoc=page,linkcolor=red,citecolor=blue]{hyperref}
\definecolor{OliveGreen}{rgb}{0,0.6,0}

\begin{document}

\title{
Solution of Wave Acceleration and Non-Hermitian Jump in Nonreciprocal Lattices}
%\textcolor{red}{Exact solution of transient dynamics in a nonreciprocal lattice}
%\\
%Non-Hermiticity-induced  wavepacket acceleration and jumps in a nonreciprocal lattice
%\\
%Unconventional wavepacket dynamics in a nonreciprocal lattice}
% Analytical description of dynamics in a nonreciprocal lattice
%Theory of dynamics in a nonreciprocal lattice

\author{Sayan Jana}\email{sayanjana@tauex.tau.ac.il}
\affiliation{School of Mechanical Engineering, Tel Aviv University, Tel Aviv 69978, Israel}
\author{Bertin Many Manda}
\affiliation{School of Mechanical Engineering, Tel Aviv University, Tel Aviv 69978, Israel}
% \affiliation{Laboratoire d’Acoustique de l’Universit\'e du Mans (LAUM),
% UMR 6613, Institut d’Acoustique - Graduate School (IA-GS), CNRS,
% Le Mans Universit\'e, Av. Olivier Messiaen, 72085 Le Mans, France}
\author{Vassos Achilleos}\email{Achilleos.Vassos@univ-lemans.fr}
\affiliation{Laboratoire d’Acoustique de l’Universit\'e du Mans (LAUM),
UMR 6613, Institut d’Acoustique - Graduate School (IA-GS), CNRS,
Le Mans Universit\'e, Av. Olivier Messiaen, 72085 Le Mans, France}
\author{Dimitrios J.  Frantzeskakis}
\affiliation{Department of Physics, National and Kapodistrian University of Athens, Athens 15
784, Greece}
\author{Lea Sirota}\email{leabeilkin@tauex.tau.ac.il}
\affiliation{School of Mechanical Engineering, Tel Aviv University, Tel Aviv 69978, Israel}

\begin{abstract}
The time evolution of initially localized wavepackets in the discrete Hatano–Nelson lattice displays a rich dynamical structure shaped by the interplay between dispersion and nonreciprocity. Our analysis reveals a characteristic evolution of the wave-packet center of mass, which undergoes an initial acceleration, subsequently slows down, and ultimately enters a regime of uniform motion, accompanied throughout by exponential amplification of the wave-packet amplitude. To capture this behavior, we develop a continuum approximation that incorporates higher-order dispersive and nonreciprocal effects and provides accurate analytical predictions across all relevant time scales. Building on this framework, we then demonstrate the existence of a 
%non-Hermitian
non-Hermiticity-induced jump --an abrupt spatial shift of the wave-packet center even in the absence of disorder-- and derive its underlying analytical foundation. The analytical predictions are 
%fully consistent 
in excellent agreement 
with direct numerical simulations of the Hatano–Nelson chain. 
%Overall, our results clarify how dispersion and nonreciprocity cooperate to produce unconventional transport behaviors, opening new possibilities for controlling wave dynamics in nonreciprocal and non-Hermitian metamaterials.
Our results elucidate the interplay between dispersion and nonreciprocity in generating unconventional transport phenomena, and pave the way for controlling wave dynamics in nonreciprocal and non-Hermitian metamaterials.
\end{abstract}

\maketitle

Non-Hermitian systems, emerging from nonreciprocal interactions, have attracted significant attention in both theoretical and experimental studies~\cite{zhu2023higher,NatRevMat2020,CalozPRapplied2018,geib2021tunable,zhang2021acoustic,DRSE}. Early research mainly focused on extending topological concepts to the non-Hermitian domain~\cite{yokomizo2019non,sato2019,ashida2020non,ninjarev22,fan24}. These investigations revealed distinctive phenomena, such as the extreme sensitivity of the spectrum to boundary conditions and the localization of eigenvectors at system edges--a behavior known as the non-Hermitian skin effect (NHSE)~\cite{OKSS2020,LTLL2023,WC2023,DSEBEC,DRSE,burst22,geib2021tunable,jana2023emerging}. Building on these foundational studies, recent work has explored the dynamical behavior of nonreciprocal lattices, uncovering potential applications including self-healing of modes~\cite{longhi_self},  invisible tunneling~\cite{jana2023tunneling}, dynamic skin effect~\cite{DSE22}, energy guiding~\cite{padlewski2024amplitude}, and also the unidirectional propagation of nonlinear excitations~\cite{VGGSMC2024,jana2025harnessing,VGBVTCC2025}.

One of the well known signatures of wave dynamics in nonreciprocal systems is their strongly unidirectional propagation along the direction of the skin modes~\cite{zhang2021acoustic,DSEninja}, where the wave is amplified along the skin-mode direction and attenuated in the opposite direction. 
In addition, nonreciprocity can induce a dynamical force that results in an additional time-dependent drift on top of the reciprocal motion. This behavior has been reported for both single-site excitations~\cite{longhi_accel,xue2024self} and finite-width wavepackets~\cite{chen2024dynamic} under weak nonreciprocity, which later cross over asymptotically to a uniform-velocity motion. Recent studies have also shown that the transition between two distinct states in nonreciprocal systems can be discrete rather than continuous in the presence of dissipative on-site disorder. This discrete transition originates from the interplay between Anderson localization~\cite{anderson1958absence} and skin-effect localization~\cite{guo2021exact}, and is termed ``non-Hermitian (NH) jump''~\cite{leventis2022non,shang2025spreading,li2025universal,kokkinakis2025dephasing}.

Wave dynamics in nonreciprocal lattices have been explored primarily through numerical studies~\cite{longhi_accel,chen2024dynamic,FLSC2025}, while analytical formulations remain not yet fully explored.
For example, the nature of time-dependent drift motion has been addressed analytically only in the weak-nonreciprocity, weakly dispersive limit, where 
%initial-time ($t \to 0$) 
analysis for short times
predicts a constant acceleration ~\cite{DSE22,lv2022curving,xue2024self}. However, a complete analysis of the drift dynamics over the full time evolution and nonreciprocity strengths remains an open question.
In particular, the dynamics in the strong non-reciprocity regime presents an opportunity for further theoretical development.
In this regime, numerical time evolution can become challenging due to instabilities and finite-precision limitations~\cite{reichel1992eigenvalues}.

In this work, we provide a comprehensive characterization of wave dynamics in discrete nonreciprocal systems arising from the interplay between dispersion and nonreciprocity. We 
%adopt an analytical strategy based on a continuum-limit expansion, 
develop an analytical approach, based on the  continuum approximation, and determine the 
%which involves solving an 
solution of the initial-value problem for the relevant partial differential equation (PDE) to elucidate wave motion in discrete lattices. This approach allows us to address several key questions: how to construct a continuum expansion that remains valid in the intermediate and strong-nonreciprocity regimes; how accurately the resulting PDE captures the underlying discrete lattice dynamics; and whether closed-form expressions for wave-packet evolution can be obtained for arbitrary nonreciprocity strength.

\begin{figure}[tb]
    \centering
\includegraphics[width=\columnwidth]{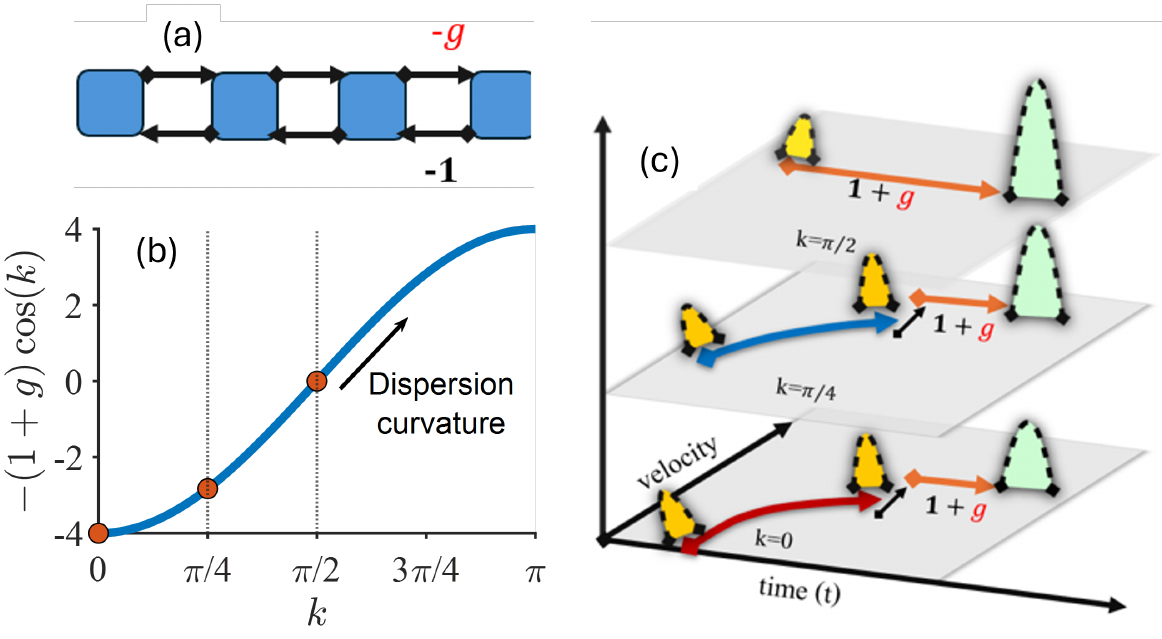}
\caption{(a) Schematic of the nonreciprocal Hatano–Nelson (HN) model. The nonreciprocity is introduced through asymmetric couplings: $g$ to the right and $1$ to the left. 
(b) Real part of the dispersion of the HN model as a function of momentum $k$. 
(c) Distinct velocity dynamics of a localized wavepacket in the Hatano–Nelson model, shown in the velocity–time plane for initial momenta $k = 0$, $\pi/4$, and $\pi/2$. 
For $k = \pi/2$, the wavepacket propagates with a constant group velocity $1+g$, exhibiting zero acceleration.
For the smaller $k$ a non-Hermitian jump is observed.
 }\label{fig1}
\end{figure}

We consider a nonreciprocal lattice system of the Hatano–Nelson~\cite{HN1998} type, shown schematically in Fig.~\ref{fig1}(a). 
Our analysis reveals that a wavepacket undergoes time-dependent acceleration by a unique power-law behavior 
---see 
%, for which 
the exact analytical form 
%is \textcolor{blue}{derived in Eq.~(\ref{phases})}. 
in Eq.~(\ref{phases}) below. 
We further analyze the time dependence of wave-packet amplitude growth, observing that both the acceleration and the amplitude growth depend strongly on the initial momentum and exhibit contrasting behaviors. In particular, as the initial momentum $k$ is tuned from the strongly dispersive regime ($k \to 0$) to the linear regime ($k \to \pi/2$)~\cite{linear_regime_note} of the dispersion relation (see its real part 
%of which is shown 
in Fig.~\ref{fig1}(b)), the initial acceleration ---as seen from the slope of the velocity--- decreases with time, while the amplitude growth is enhanced. 

We find that the growth rate is maximal for wavepackets initialized near $k \to \pi/2$, as schematically illustrated in Fig.~\ref{fig1}(c). The momentum ($k$)-dependent growth reveals an intriguing dynamical feature, namely a ``non-Hermitian'' jump occurring in a disorder-free nonreciprocal lattice. We analytically demonstrate that, rather than a continuous evolution, the wave motion can exhibit a discontinuous jump between two distinct velocity regimes in time, controlled by the nonreciprocity strength and the spatial width of the wavepacket.

\begin{figure*}[tb]
    \centering    \includegraphics[width=0.95\textwidth,height=0.40\textheight]{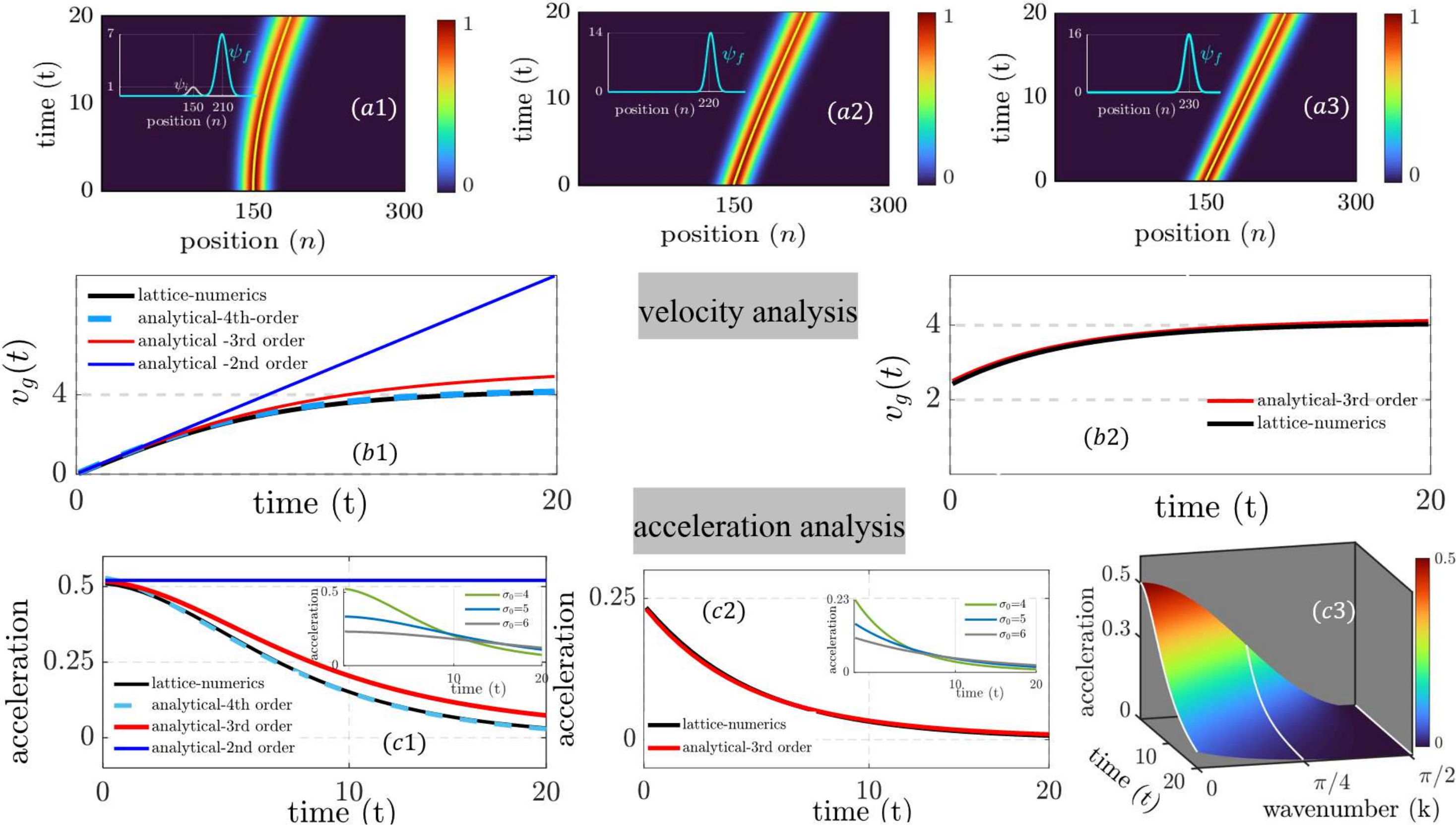}
    \caption{ Normalized spatiotemporal response $|\Psi_{n}(t)|$ of the initial wavepacket $\Psi_{\rm in}$ in the Hatano–Nelson chain, as given by Eq. (\ref{initialwavepacket}): Panels (a1), (b1), and (c1) correspond to initial momenta $k_i = 0$, $\pi/4$, and $\pi/2$, respectively. The inset in (a1) shows the initial profile $\Psi_{\rm in}$ at $t = 0$ and the time-evolved wavepacket $\Psi_f$ at $t = t_f$. The corresponding final profiles for $k_i = \pi/4$ and $\pi/2$ are shown in the insets of (b1) and (c1). The progression from (a1) to (c1) highlights that the amplitude of $|\Psi_f|$ is largest for $k_i = \pi/2$. Group velocity and acceleration from lattice simulations and PDE theory: Panels (b1) and (c1) correspond to $k_i=0$, panels (b2) and (c2) to $k_i=\pi/4$, and panel (c3) shows the acceleration phase diagram in the $(t, k)$ plane. In all the panels we chose $\sigma_0=4$, $g=3$.}
    \label{fig2}
\end{figure*}
\textit{\textcolor{black}{Time dependent acceleration}}. We consider a nonreciprocal lattice governed by the dimensionless Schr\"odinger equation:
\begin{equation}
-\mathrm{i} \frac{d\Psi_n}{dt} = g\Psi_{n-1} + \Psi_{n+1},
\label{schrodinger}
\end{equation}
where $\Psi_n(t)$ denotes the field amplitude of the wave at time $t$ and site $n$ ($n = 1, 2, \dots, N$), and $g > 1$ quantifies the nonreciprocity in the couplings. We assume that \eqref{schrodinger} is supplemented with an initial condition $\Psi_{n}(0)$ in the form of a 
%an initial 
Gaussian wavepacket, with momentum $k_i$, %that is created around 
located at the site $n_{0}$ of the lattice %in 
(see Fig. \ref{fig1}(a)), namely
%given by
\begin{align}
\Psi_{n}(0) = \frac{1}{(2\pi \sigma_0^2)^{1/4}} e^{\mathrm{i} k_i n}
e^{-\frac{(n- n_0)^2}{4\sigma_0^2}},
\label{initialwavepacket}
\end{align}
where $\sigma_0$ denotes the initial spatial width of the wavepacket. 
%We assume that the time evolution of the 
 %initial 
%wavepacket is governed by 
%solving 
%the dimensionless Schr\"odinger equation
%\begin{equation}
%-\mathrm{i} \frac{d\Psi_n}{dt} = g\Psi_{n-1} + \Psi_{n+1},
%\label{schrodinger}
%\end{equation}
%where $\Psi_n(t)$ denotes the field amplitude of the wave at time $t$ and site $n$ ($n = 1, 2, \dots, N$), and $g > 1$ quantifies the nonreciprocity in the couplings. 
%Eq.~
While the initial value problem 
(\ref{schrodinger}),\eqref{initialwavepacket}  can be solved numerically, 
%using the initial condition $\Psi_{\mathrm{in}}$ from Eq.~(\ref{initialwavepacket}), providing the time-evolved wave function $\Psi_n(t)$ which contains the information of the nonreciprocal dynamics. 
%
our aim is to gain analytic insights into the dynamics. 
%governed by 
For this purpose, we derive from 
Eq.~(\ref{schrodinger})
%, we derive 
an effective continuum PDE, %that is 
first order in time $t$ and higher order in space $x$, which is valid near a chosen wavenumber $k_i$ of the dispersion. With the discrete field variable $\Psi_n(t)$ replaced by its continuum counterpart $\Psi(x,t)$, the resulting PDE reads
%takes the following form:
\begin{subequations}\label{pde}
\begin{eqnarray}
&&\mathrm{i}\frac{\partial \Psi(x,t)}{\partial t} = \mathcal{H}\Psi(x,t), \label{pde:a} \\
&&\mathcal{H} = 
\sum_{m=0}^{4} c_m(k_i,g)\frac{\partial^m}{\partial x^m}
%c_0 
%    + c_1 \frac{\partial}{\partial x}
 %   + c_2 \frac{\partial^2}{\partial x^2}
  %  + c_3 \frac{\partial^3}{\partial x^3}
   % + c_4 \frac{\partial^4}{\partial x^4}, 
   \label{pde:b}
\end{eqnarray}
\end{subequations}
where the expansion coefficients $c_0, c_1, c_2, c_3, c_4$ are functions of $(k_i, g)$. Consequently, for each initial momentum $k_i$, a separate PDE must be solved. The details of the derivation are provided in the Supplemental Material~\cite{supplementary}.
Starting from the initial condition $\Psi_{\mathrm{in}}^{}(x,0)$ in Eq.~(\ref{initialwavepacket}), the time-evolved wavepacket reads
\begin{equation}
\Psi(x,t) = e^{-\mathrm{i} \mathcal{H}^{} t}\,\Psi_0(x)
%{\mathrm{in}}^{}(x,0),
\label{Lea}
\end{equation}
where 
$\Psi_0$ is the continuum counterpart of \eqref{initialwavepacket}. 
Substituting $k$$\to$ $-\mathrm{i}\partial_{x}$, \eqref{Lea} can be expressed in the following integral form:
\begin{align}\label{integral}
\Psi(x,t) &= \int_{-\infty}^{\infty} 
e^{\mathrm{i} k x}e^{-\mathrm{i}\big(c_0+c_1 k + c_2 k^2 + c_3 k^3 + c_4 k^4\big)t }
\widehat{\Psi}_0(k)\, \mathrm{d} k, \nonumber\\
&= \int_{-\infty}^{\infty}e^{\phi(k,x,t)} \mathrm{d} k,
\end{align}
where $\widehat{\Psi}_0$ is the Fourier transform of $\Psi_0$.
%Near $k_i\to 0$, for the weakly dispersive case around the i.e.($c_3,c_4\to 0$), Now if we try to sove Eq.\ref{eq4}  %is the moving properties of the pulse center of mass. 
% We apply the spectral method for solving arbitrary PDE~\cite{PPWC2003}.Consequently, we look for a solution $y(x,t)=\int u_p(t)e^{ipx}dp$ which are a superposition of periodic functions of variable $p$.
Our task is to 
%solve 
evaluate the integral in Eq.~(\ref{integral}). With higher-order dispersive terms ($c_3, c_4 \neq 0$), the integral cannot be solved exactly in a straightforward manner, unlike in the weakly dispersive limit where these terms are negligible.
%To solve Eq.~(\ref{integral}), 
We thus apply the saddle-point method 
%(also known as the stationary-phase or steepest-descent method)
~\cite{wong2001asymptotic} to approximate the integral. This is obtained by considering that
the leading behaviors come from the contributions of the stationary points $k_s$ of the exponent of the integrand, i.e., $\partial\phi/\partial k|_{k_s}=0$, which leads to 
%It follows that,
$\Psi(x,t)\approx e^{\phi \left(k_s\left(x,t\right)\right)}$ (see~\cite{supplementary}).
The time-evolved wavepacket can then be approximated as
\begin{align}\label{xcompsi}
\Psi(x,t) &\approx e^{\phi\big(k_s(x,t)\big)} \propto e^{-\big(x - x_{\rm com}(t)\big)^2},
\end{align}
where $x_{\rm com}(t)$ denotes the center-of-mass (COM)
%position 
of the wavepacket at time $t$.
For $k_i = 0$, considering a moderate dispersive case with ($c_3 \neq 0$, $c_4=0$), we obtain an analytical expression for the time dependence of the COM,
\begin{equation}\label{xcom}
x_{\rm com}(t) \approx \frac{2 + \gamma^2}{6\gamma} \left( \sqrt{36\sigma_0^4 + g \gamma^2 (6 + \gamma^2) t^2} - 6\sigma_0^2 \right),
\end{equation}
where $\gamma = \log(g)$.
Consequently, a tractable expression for the acceleration can be obtained from $a_{\rm com}(t)=\ddot{x}_{\rm com}(t)$,
\begin{equation}\label{acom}
a_{\rm com}(t) \approx \frac{2\sqrt{3} \,\sigma_0^4 \, g \gamma (2 + \gamma^2) (6 + \gamma^2)}{\left[12\sigma_0^4 + g \gamma^2 (6 + \gamma^2) t^2 \right]^{3/2}}.
\end{equation}
Equation~(\ref{xcom}) provides %important 
essential information about the interplay between the wavepacket width $\sigma_0$, the nonreciprocity parameter $g$, and time $t$, enabling a detailed analysis of the dynamics. Moreover, it highlights a significant difference from the weakly dispersive, weakly nonreciprocal case ($c_3, c_4 \to 0$)~\cite{DSE22,lv2022curving}. In this limit, the integral in Eq.~(\ref{integral}) can be solved exactly~\cite{CDL1986,S2016}, giving
\begin{equation}\label{xcom_weakly}
x_{\rm com}(t) = a_{\rm com} t^2,
\end{equation}
with constant acceleration $a_{\rm com} = \frac{2\gamma g}{\sigma_0^2} (1+\frac{\gamma^3}{6}+\frac{\gamma^4}{192})$. 
In contrast to~Eq. (\ref{xcom_weakly}), Eq.~(\ref{acom}), obtained in the moderately strong dispersive limit, shows that the acceleration of the wavepacket is time-dependent and follows a nontrivial power-law behavior in $t$, as discussed in detail as follows.

To analyze the time dependence encoded in Eqs.~(\ref{xcom}) and~(\ref{acom}) in more detail, we compare them with the numerical solution of 
%the time-dependent 
Eq.~(\ref{schrodinger}). 
The spatiotemporal evolution obtained from lattice simulation is shown in Fig.~\ref{fig2}(a1)~\cite{HNW1993,DMMS2019}. Although the initial group velocity $v_g$ is zero, the center of mass $x_{\rm com}(t)$ begins to move, acquiring an additional time-dependent drift. We plot in Fig.~\ref{fig2}(b1) the group velocity $v_g(t) = \dot{x}_{\rm com}(t)$ obtained from the numerical simulation and from various approximations of the PDE
%in Eq.~
(\ref{pde}). The lattice simulation (black) matches well with the solution of the strongly dispersive %fourth 
4th-order PDE ($c_4 \neq 0$) shown in blue. However, as we gradually consider lower-order approximations—namely the 
%third (
3rd-order PDE ($c_3 \neq 0$, $c_4 = 0$, red) and the 
%second (
2nd-order PDE ($c_1, c_2 \neq 0$, $c_3 = c_4 = 0$, purple)—the solutions show increasing deviation from the numerical lattice simulation.

The corresponding acceleration extracted from the simulation is plotted in black in Fig.~\ref{fig2}(c1) and matches exactly with the solution of the strongly dispersive 4th-order PDE shown in blue. As we gradually consider lower-order approximations, the 3rd-order PDE obtained from Eq.~(\ref{acom}) (red) shows a small deviation from the lattice simulation, but qualitatively captures the time dependence of the acceleration. In contrast, the 2nd-order PDE solution given in Eq.~(\ref{xcom_weakly}) exhibits a constant acceleration over the entire time frame.
Next, we analyze the time dependence of the acceleration in Eq.~(\ref{acom}) and distinguish three stages in the evolution of a moving wavepacket with respect to the characteristic time
\begin{equation}
T^{\star} = \frac{\sigma_0^2}{\gamma \sqrt{g\left(6 + \gamma^2\right)}}.
\end{equation}
The first stage happens at small times $t\ll T^{\star}$, in which we find that the acceleration $a_{com}(t)\propto g \gamma \sigma_0^{-2}$ is constant, as seen in the weakly dispersive regime.
The duration of the constant-acceleration stage increases as the initial pulse width $\sigma_0$ becomes larger and/or the nonreciprocal strength decreases. %(see $T_{\rm{weak}}$).

The second stage starts when $t\gtrsim T^\star$, leading to the time dependent part of the acceleration to dominate.
It follows that $a_{\rm com}\propto \sigma^4 g^{-1/2} \gamma^{-2} t^{-3}$, 
i.e. the acceleration decays with time following this specific power law.
Consequently, it leads to an overall slowdown of the wavepacket motion.
Ultimately after a long time, $t\gg T^{\star}$, we enter the third stage 
%in which the 
where $a_{com}=0$.
%{\it The key distinction of the wave-packet dynamics in a strongly dispersive medium, such as the HN lattice, is that the non-Hermitian wave acceleration is a transient feature occurring at finite times.}
We therefore summarize the expected stages of acceleration in the wave dynamics as
\begin{equation}\label{phases}
a_{\rm com}(t) \propto 
\begin{cases}
\dfrac{2g \gamma}{\sigma_0^2}, & t < T^{\star} \quad \text{(constant-acceleration)} \\ 
\dfrac{\sigma_0^4}{g^{\frac{1}{2}} \gamma^2 t^{3}}, & t > T^{\star} \quad \text{(de-acceleration)} \\ 
0, & t \gg T^{\star} \quad \text{(uniform motion)}
\end{cases}
\end{equation}
For a fixed value of nonreciprocity $g$, all these stages can be observed in the inset of Fig.~\ref{fig2}(c1) with varying $\sigma_0$. When $\sigma_0$ is small ($\sigma_0 = 4$), $T^{\star}$ is also small, so the de-acceleration phase dominates over the constant acceleration phase. As $\sigma_0$ gradually increases, $T^{\star}$ increases, allowing the constant acceleration phase to gradually extend over a longer time span, as seen in the plots for $\sigma_0 = 5$ (blue) and $\sigma_0 = 6$ (gray). 
Similarly, we analyze the time evolution of an initial wavepacket $\Psi(x,0)$ with a finite momentum $k_i \neq 0$, specifically around $k_i = \pi/4$. In this case, the coefficients $c_1, c_2, c_3$ are extracted from the dispersion relation by expanding the PDE in Eq.~(\ref{pde}) about $k_i = \pi/4$~\cite{supplementary}. Solving Eq.~(\ref{xcom}) yields the center-of-mass trajectory
\begin{equation} \label{eqboomerang}
    x_{\mathrm{com}}(t) = f(k_i, \sigma_0, g).
\end{equation}
The explicit dependence of Eq.~(\ref{eqboomerang}) on the wave-packet width $\sigma_0$ and the nonreciprocity $g$ is given in Ref.~\cite{supplementary}. The prediction of Eq.~(\ref{eqboomerang}) is in excellent agreement with the lattice simulation, as shown in Fig.~\ref{fig2}(b2)–(c2).
%
%The spatiotemporal evolution obtained from lattice simulations is shown in Fig.~\ref{fig2}(a1). In %Figs.~\ref{fig2}(b2) and~\ref{fig2}(c2), we plot the group velocity $v_g(t)$ and the acceleration %$a(t)$ obtained from Eq.~(\ref{eqboomerang}) (red), together with the lattice simulation results %(black). The excellent agreement between the analytical expressions and the numerical simulations again %highlights the importance of retaining spatial terms up to third order in the PDE formulation. The %inset of Fig.~\ref{fig2}(c1) shows the dependence of the acceleration on the initial width $\sigma_0$, %revealing a gradual decrease in acceleration as the wavepacket becomes broader.

%.% This matches the numerical result in Fig.~\ref{fig2}(c1) and corresponds to the zero-acceleration curve in Fig.~2(e).
%To visualize the variation of the wave-packet acceleration at the $j^{\mathrm{th}}$ wave vectors $k_i^j$, spanning from the strongly dispersive region ($k_i = 0$) to the linear regime ($k_i = \pi/2$) in the energy dispersion, as a function of time $t$, we consider Eq.~(\ref{xcompsi}).
To visualize how the acceleration varies with wave vector $k_i$ (ranging from the dispersive region, $k_i = 0$, to the linear regime, ($k_i = \pi/2$) as a function of time $t$, we consider Eq.~(\ref{xcompsi}). For each $k_i$, the coefficients $c_1, c_2, c_3$ are computed to construct the phase diagram in the $(t, k = k_i)$ plane, shown in Fig.~\ref{fig2}(c3). This diagram reveals a continuous reduction of acceleration as one moves from the dispersive regime ($k_i\to 0$) toward the nearly linear region of the dispersion. This behavior directly reflects the dependence of $a_{\rm com}(t)$ on the local curvature of the band dispersion.
When $k = \pi/2$, the dispersion becomes effectively linear, and a 2nd-order truncation of the PDE in Eq.~(\ref{pde}) is sufficient, with $c_1 = -(1+g)+\mathrm{i}(1-g)\pi/2$ and $c_2 =(1-g)/2$, leading to a constant group velocity $v_g = 1 + g$.

\begin{figure}[tb]  
    \centering
    \includegraphics[width=.85\columnwidth]{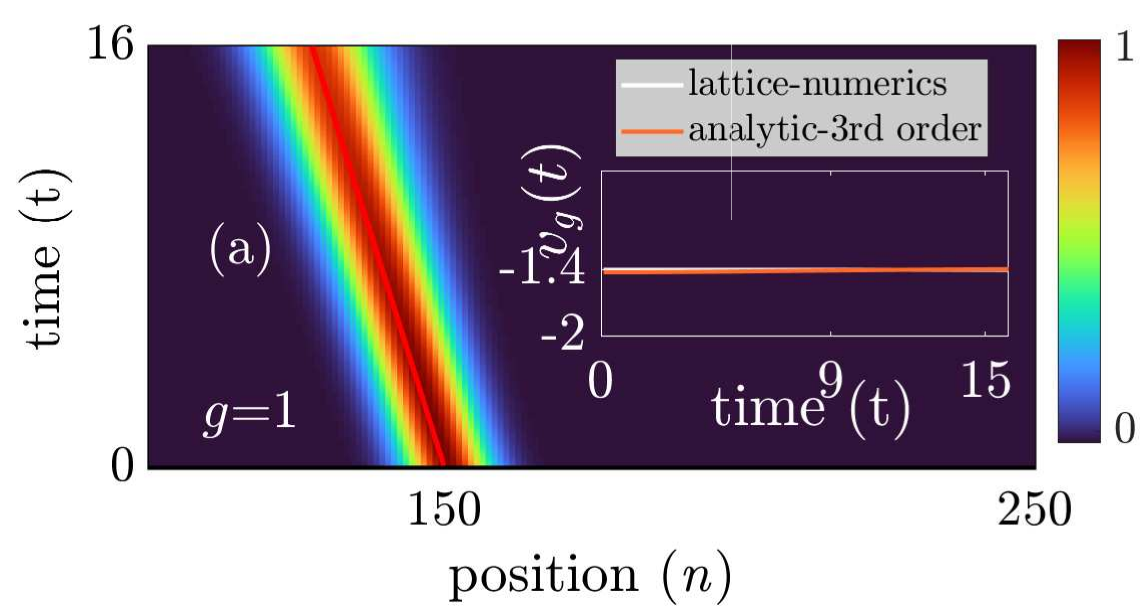} \\
    \includegraphics[width=.85\columnwidth]{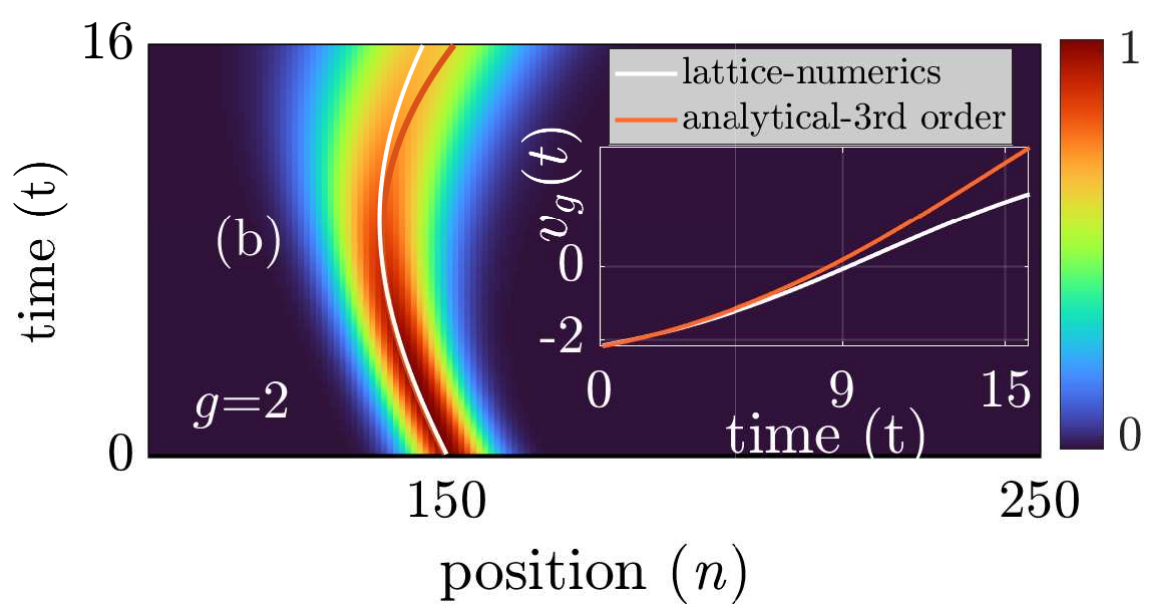}
    \caption{Non-Hermitian Boomerang Effect: (a) Spatiotemporal response of the wave dynamics under Hermitian ($g = 1$) evolution for $k_i = -\pi/4$. 
(b) Nonreciprocal ($g = 2$) dynamics for the same initial momentum, $k_i = -\pi/4$, revealing a boomerang-like reversal in the wave's motion. The insets in (a) and (b) shows the wave-packet group velocity $v_g(t)$ obtained from simulations of Eq.(~\ref{schrodinger}) (white) and the analytical expression in Eq.(~\ref{eqboomerang}) (orange). Specifically, $v_g(t)$ in (b) show a sign reversal of the group velocity from negative to positive over time. For both panels, we choose $\sigma_0 = 3.5$.
 }
 \label{fig3}
\end{figure}

An interesting signature of the time dependent accelerated motion appears in the direction reversal of the wavepacket, as shown in Fig.~\ref{fig3}. We start with a wavepacket at $k_i = -\pi/4$ in the Hermitian chain ($g = 1$). As expected, due to the negative group velocity, the wavepacket moves to the left with a constant group velocity of $v_g \approx -1.4$, as shown in panel (a) and the inset therein.

However, once we switch on nonreciprocity ($g = 2$), the dynamics changes. Initially, the wavepacket still tends to move to the left, dictated by the negative group velocity, but as time passes, the additional time-dependent nonreciprocal drive, as shown in Eq.~(\ref{eqboomerang}), begins to act (see~\cite{supplementary}). Since this nonreciprocal force (acceleration) contributes to a positive group velocity, it gradually counteracts the initial motion, and with increasing time, it 
%eventually 
reaches a particular moment 
at which it fully balances the initial motion, resulting in a zero group velocity ($v_g = 0$). %The dependence of $t_b$ on $g$ and $\eta$ also can be seen from Eq.~(\ref{eqboomerang}). 
Beyond this point ($t \approx 9$), the nonreciprocal contribution dominates and pushes the wavepacket toward the wave-amplifying direction, leading to a complete reversal of motion. We refer to this phenomenon as the \textit{Non-Hermitian Boomerang Effect}, illustrated in Fig.~\ref{fig3}(b). The sign reversal of the group velocity $v_g(t)$, obtained from numerical simulations of Eq.~(\ref{schrodinger}) (white) and from the corresponding analytical expression in Eq.~(\ref{eqboomerang}) (orange), is shown in the inset of Fig.~\ref{fig3}(b).
%
%The sign reversal of the group velocity, obtained from the numerical simulation of Eq.~\ref{schrodinger} (white) and the corresponding analytical expression (orange), is shown in the inset of Fig.~\ref{fig3}(b). We observe that the analytical formula in Eq.~(\ref{eqboomerang}) accurately predicts the reversal time. %Beyond $t_b$, however, the analytical curve in panel (b) shows a slight deviation from the numerical result. This is expected, as the expression was derived around $k_i = -\pi/4$, and once the wavepacket reaches the point where $v_g = 0$, corresponding to a wavenumber $k$ approaching $k(t) \to 0$, this approximation is no longer sufficient to fully capture the subsequent dynamics. Nevertheless, the analytical formulation still correctly captures the sign change of the group velocity.

\textit{Non-Hermitian Jump}. We observe from the lattice simulations that not only the acceleration but also the amplitude growth depends on the initial momentum $k_i$. To quantify this effect, we introduce the parameter $\beta = \log_{10} |\Psi(t_f)|$, where $t_f$ denotes the final evolution time used in the simulation, and plot the normalized amplitude. 
%as $|\Psi(t_f)| = \beta |\Psi(t_f^{})^{\nu}|$, with $\Psi(t_f)^{\nu}$ normalized under area conservation. 
This behavior is visible in the insets of Figs.~\ref{fig2}(a)–(c), where $|\Psi(t_f)|$ is shown for gradually increasing initial momenta $k_i = 0$, $\pi/4$, and $\pi/2$, respectively. We find that the amplitude growth is largest for $k_i = \pi/2$. We analytically obtained (see~\cite{supplementary}) that, for a relatively broad wavepacket, the time dependence of the amplitude growth is proportional to $\lvert \Psi(x,t) \rvert \propto e^{\theta_{k_i}(t)}$. For an initial wavepacket centered at $k_i =0$ the $\theta_{k_i}(t)$ evolves as
\begin{equation} \label{amp_k_0}
    \theta_{k_i=0}(t) = \frac{(g-1)^2 t^2}{2\sigma_{0}^2} \, .
\end{equation}
In contrast, for a $k_i = \pi/2$ wavepacket, $\theta_{k_i}(t)$ follows
\begin{equation} \label{amp_k_pi_2}
    \theta_{k_i=\pi/2}(t) \approx \frac{\pi^2}{4} (g-1)\, t \, .
\end{equation}
Comparing Eqs.~(\ref{amp_k_0}) and~(\ref{amp_k_pi_2}) shows that the presence of $\sigma_0$ suppresses the growth rate of $\theta_{k_i=0}(t)$ relative to $\theta_{k=\pi/2}(t)$. Consequently, the amplitude growth is largest for $k_i = \pi/2$.

\begin{figure}[tb]  \label{fig:single_column}
    \centering
    \includegraphics[width=.99\columnwidth]{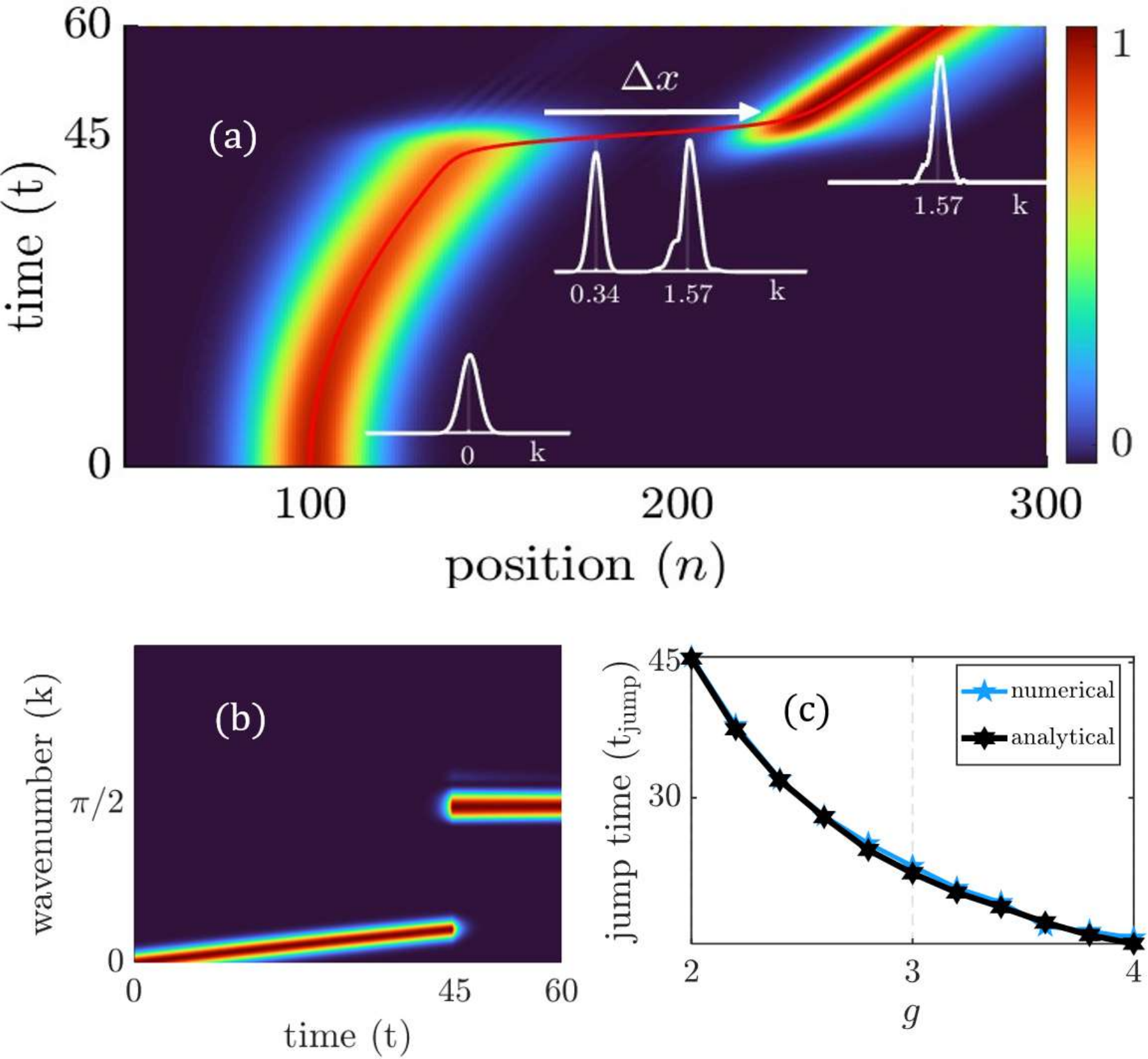} 
    \caption{ Non-Hermitian Jump: (a) Normalized spatiotemporal evolution $|\Psi_n(t)|$ for $g$=2 of the initial wavepacket $\Psi_{\rm in}$, with the corresponding momentum-space distribution $|\Psi_k(t)|$ shown at $t = 0$, $45$, and $60$. 
    (b) Time dependence of the mode weights $\varphi_k(t)$ during the evolution. 
    (c) Jump time $t_{\mathrm{jump}}$ as a function of nonreciprocity $g$ for $\delta = 45$. 
    All panels correspond to $k_i = 0$, and $\sigma_0 = 8$. 
 }
 \label{fig4}
 \end{figure}

We next show that this momentum dependence of the amplitude growth leads to an intriguing phenomenon. We start with a wavepacket centered at $k_i = 0$, with width $\sigma_0 = 8$ and nonreciprocity $g = 2$. The corresponding time evolution is shown in Fig.~\ref{fig4}(a). We find that, although the trajectory initially follows Eq.~(\ref{xcom}), after a critical time $t_\mathrm{jump}$ the dynamics, $\Psi_n (t)$ in real space undergo a discontinuous transition: a second wavepacket emerges, propagating with the constant velocity $1+g$. 

The Fourier spectrum of the time-evolved wavepacket at successive time instants, 
%\textcolor{blue}{
as shown in the inset of Fig.~\ref{fig4}(a), reveals important details. At the jump point, an additional peak suddenly emerges near $k = \pi/2$, alongside the original peak that follows Eq.~(\ref{xcom}). Beyond the critical time $t_\mathrm{jump}$, the newly emerged wavepacket dominates the dynamics and propagates with momentum $k = \pi/2$, as confirmed by the Fourier spectrum.  
This phenomenon represents a peculiar dynamical feature, unexpected in disorder-free nonreciprocal systems: an effective jump of the dominant momentum component from $k = 0$ to $k = \pi/2$, is termed ``non-Hermitian jump''.

The key point of this phenomenon is that an initial excitation 
%within the bulk 
$\Psi_n (0) = \sum_k \varphi_k (0) u_{k, n}$, centered at wave number $k_i$, will activate all eigenmodes $u_{k,n} = e^{\mathrm{i}kn}$ of amplitudes $\varphi_k(0)$, corresponding to the eigenenergies
\begin{equation} \label{disp}
    E_k = -(g+1)\cos k + \mathrm{i}(g-1)\sin k .
\end{equation}% $E_k = -(g+1)\cos k + \mathrm{i}(g-1)\sin k$.
The amplitude $\varphi_k(0)$ is maximum at $k_i$ and decays as $k$ moves away from this value.
%We find that the modal decomposition of 
Decomposing the wavefunction 
%\eqref{schrodinger} 
as $\Psi_n(t) = \sum_k \varphi_k(t) u_{k,n}$, %where $\varphi_{k}(t) = \varphi_{k}(0)e^{-\mathrm{i}E_kt}$,
%Thus, 
we find that, the equations of motion Eq.~\eqref{schrodinger} of the discrete Hatano-Nelson lattice, in the normal mode space reduces to $\mathrm{i}\dot{\varphi}_k = E_k \varphi_k$.
It follows that the modal amplitudes 
\begin{equation} \label{mode_amp_k}
    \varphi_{k}(t) = \varphi_{k}(0)e^{-\mathrm{i}E_kt} = \varphi_{k}(0)e^{t(g-1)\sin k}\, e^{-\mathrm{i}t(g+1)\cos k}.
\end{equation}
% which grows exponentially in time. 
Clearly, $\varphi_{k}(t)$ grows in time following a $k$-dependent growth rate $t(g-1)\sin k$ which is minimal at $k=0$ and increases as $k\rightarrow \pi/2$, where it becomes maximal.
Consequently, irrespective of the distribution of $\varphi_k(0)$ there exists a time $t$ for which $\varphi_{k=\pi/2}(t)$ becomes larger than $\varphi_{k_i}(t)$, leading to a wave jump at finite time, also in the mode representation.
Figure~\ref{fig4}(b) displays the time dependence of the numerically obtained $\varphi_k (t)$ ($k=2\pi l/N$) using the real-space data of Fig.~\ref{fig4}(a)~\cite{advanpix}.
It follows that such a scenario becomes clearly visible, in both the real and the mode domains, the latter being comparable to the standard Fourier transform shown in the insets in Fig.~\ref{fig4}(a).
% Interestingly these amplitude growth rates $t(g-1)\sin k_i$ for $k_i=0$ and $k_i = \pi/2$ are different when in mode and real space representations as per Eq.~\eqref{amp_k_0} and Eq.~(\ref{amp_k_pi_2}).

We focus on the real space representation to obtain an analytical estimate for the jump time $t_{\mathrm{jump}}$.
Bearing the above in mind, we consider the time evolution of the wavepacket amplitude approximates as
\begin{equation}
\Psi(x, t) \propto e^{\theta_{k_i=0}(t)} + e^{-\delta} e^{\theta_{k_i=\pi/2}(t)}.
\end{equation}
The physical assumption is that, although the wavepacket is initially prepared around $k_i = 0$, its finite width $\sigma_0$ implies a weak excitation of other momenta, including $k = \pi/2$. 
Since the wavepacket is centered at $k_i = 0$, the initial weight of the $k = \pi/2$ component is exponentially small and is parametrized by the factor $e^{-\delta}$, with $\delta \gg 1$.
Due to the different time dependences of the growth rates $\theta_{k_i}(t)$, the initially dominant $k_i = 0$ component is eventually overtaken by the $k_i = \pi/2$ component, and the jump time $t_{\mathrm{jump}}$ is determined by the condition $e^{\theta_{k_i=0}(t_{\mathrm{jump}})} \approx e^{-\delta} e^{\theta_{k_i=\pi/2}(t_{\mathrm{jump}})}$, yielding
\begin{equation} \label{t_cric}
t_{\mathrm{jump}} =\frac{\sigma_{0}^2}{4(g-1)}\left(\pi^2 - 2\sqrt{\sigma_{0}^2\pi^4 - 32\delta}\right).
\end{equation}
% Note that a similar estimate of the jump time can also be found using the mode space variables, $\varphi_k$ [Eq.~\eqref{mode_amp_k}].
% We numerically compute $t_{\mathrm{jump}}$ using the $\Psi_n (t)$-data obtained through lattice simulations and compared it with Eq.~(\ref{t_cric}).
% Fig.~\ref{fig4}(c) shows the result of this comparison, depicting a good agreement between the theory and numerical simulations.
Note that a similar estimate of the jump time can also be obtained using the mode-space variables, $\varphi_k$, Eq.~\eqref{mode_amp_k}.
We numerically extract $t_{\mathrm{jump}}$ from the numerical lattice dynamics $\Psi_n(t)$ and compare it with the theoretical prediction given by Eq.~\eqref{t_cric}.
Figure~\ref{fig4}(c) presents this comparison and demonstrates good agreement between theory and simulations.

In conclusion, we analytically and numerically investigated the time evolution of 
%initially localized 
wavepackets in the discrete Hatano–Nelson lattice and clarified how nonreciprocity and dispersion jointly shape their dynamical behaviors.
Using a continuum approximation of the lattice,  
leading to a 3rd-order PDE,
%with a PDE truncated at third-order spatial derivatives, 
%extracted from the lattice, 
we showed that the dynamics of the wavepacket's COM 
%center of mass  
(initially starting with zero group velocity) exhibits three distinct stages: 
%namely 
(i) an early-time acceleration, followed by (ii) a deceleration that ultimately settles into (iii) uniform motion 
%, as given in 
(see Eq.~(\ref{phases})).
This COM motion is accompanied by a wave-packet amplitude that grows exponentially in time. %\textcolor{blue}{
We analytically showed that both the acceleration and amplitude growth depend explicitly on the initial momentum, giving rise to a boomerang-like reversal of the group velocity.
%}

We further identified a non-Hermitian jump, corresponding to an abrupt shift of the wave-packet COM during its evolution. We provided an explicit analytical description of this phenomenon, capturing its dependence on nonreciprocity and the initial wave-packet width, Eq.~(\ref{t_cric}), and confirming that it can occur even without disorder.
These effects have no analogue in Hermitian systems and are expected to be relevant for controlling waves in nonreciprocal and non-Hermitian metamaterials.

\begin{acknowledgments}
   \textit{
Funded by the European Union. Views and opinions expressed are however those of the author(s) only and do not necessarily reflect those of the European Union or the European Research Council Executive Agency. Neither the European Union nor the granting authority can be held responsible for them. V.A was supported by ERC grant NASA -  101077954.
S.J., B.M.M., and L.S. were supported in part by the Israel Science Foundation Grants No. 2177/23 and 2876/23.}
\end{acknowledgments}

\paragraph{Note added.} After completion of this article, we became aware of the recent work by He and Ozawa~\cite{HO2025}.
In their study, the authors investigate the spreading dynamics of initially localized wave packets in Hatano–Nelson type models.
Using a 1st order approximation in momentum space, they predict the existence of non-Hermitian wave-packet jumps in spatially periodic nonreciprocal lattices and characterize their main features.
%As such, their theoretical analysis is mainly valid in the regimes of weak dispersion and/or weak nonreciprocity.
In the present work, the theoretical analysis is carried out in both real and momentum spaces, considering terms in the dispersion up to the 3rd and 4th orders.
Thus our analysis is broadband, and extend to a large range of nonreciprocal strengths.
%Nevertheless, the work of He and Ozawa, and the present study strongly corroborate each other.

\bibliography{nonlinear}

\end{document}